\documentstyle[preprint,eclepsf]{jpsj}

\def\runtitle{On the Spin Gap Phase in
$\lambda$-(BETS)$_2$GaX$_z$Y$_{4-z}$}
\def\runauthor{Hitoshi {\sc Seo} and Hidetoshi {\sc Fukuyama}}

\def\lsim{\lower -0.3ex \hbox{$<$} \kern -0.75em \lower 0.7ex \hbox{$\sim$}}
\def\gsim{\lower -0.3ex \hbox{$>$} \kern -0.75em \lower 0.7ex \hbox{$\sim$}}

\title{On the Spin Gap Phase in \\
$\lambda$-(BETS)$_2$GaX$_z$Y$_{4-z}$}
\author{Hitoshi {\sc Seo}\footnote{E-mail: hseo@watson.phys.s.u-tokyo.ac.jp} 
and Hidetoshi {\sc Fukuyama}}
\inst{Department of Physics, Faculty of Science, University of Tokyo, Tokyo 113}

\recdate{}

\abst{The nature of the insulating ground state 
in quasi-two-dimensional 
organic conductor $\lambda$-(BETS)$_2$GaX$_z$Y$_{4-z}$, 
where the existence of a spin gap 
is suggested by susceptibility measurement, has been 
studied theoretically. 
Hartree-Fock calculations at absolute zero temperature
show that if the on-site Coulomb interaction exceeds some critical value, 
then antiferromagnetic spin ordering emerges and eventually leads to 
an insulating state which can be considered as a two-dimensional 
localized spin system. 
Based on the quantum Monte Carlo simulations of Katoh and Imada, 
we will show that this system locates near the boundary 
between the antiferromagnetic 
phase and the spin gap phase, 
so that experimental facts can be explained.}
\kword{BETS, spin gap, Hartree-Fock approximation, on-site Coulomb interaction,
antiferromagnetic spin ordering}

\begin{document}
\sloppy
\maketitle

\begin{figure}
\begin{center}
\epsfile{file=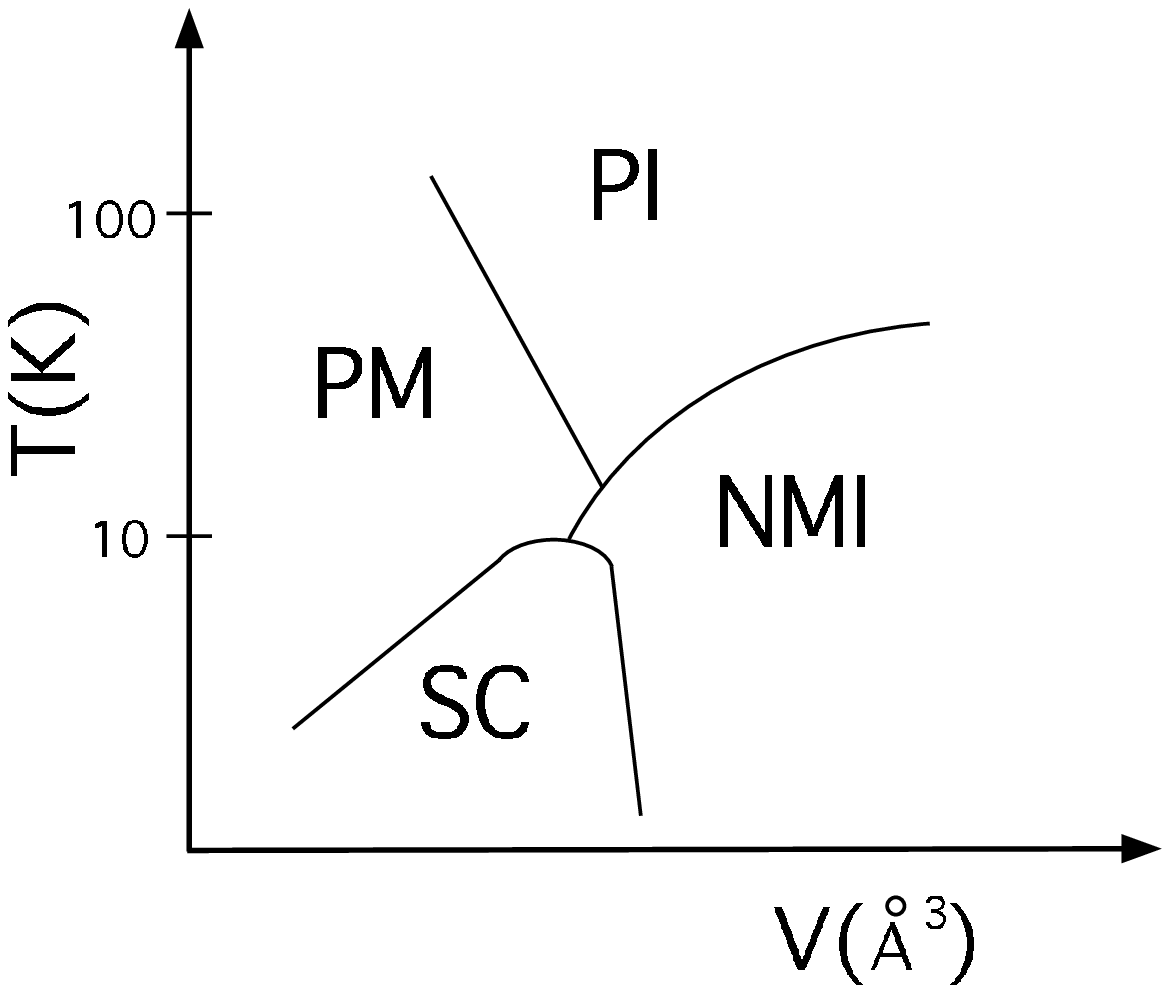,height=5.5cm}
\end{center}
\caption{Experimental phase diagram of $\lambda$-(BETS)$_2$GaX$_z$Y$_{4-z}$ 
on the plane of unit cell volume ($V$) and temperature ($T$).\cite{Koba} 
NMI, SC, PI and PM stands for the nonmagnetic insulating, superconducting, 
paramagnetic insulating and paramagnetic metallic phases, respectively.}
\label{phase}
\end{figure}

Organic conductors made of BEDT-TSeF (bisethylenedithiotetraselenafulvalene, 
abbreviated as BETS), 
which is a seleno-analog of the well-known BEDT-TTF (abbreviated as ET),  
have two-dimensional (2D) character as ET compounds does. 
Among them, a new series of superconductors 
$\lambda$-(BETS)$_2$GaX$_z$Y$_{4-z}$ (X,Y=F,Cl,Br) 
exhibit interesting 
electronic properties. 
By recent resistivity and magnetic susceptibility measurements
with different $z$-values and different pressures, 
the phase diagram of $\lambda$-(BETS)$_2$GaX$_z$Y$_{4-z}$ 
was proposed\cite{Koba} as shown in Fig. \ref{phase}. 
The substitution of X and Y to smaller atoms, 
the decrease of the $z$-value when X is a larger atom than Y 
or condition of applied pressure 
corresponds to the reduction of the unit cell volume $V$. 
Since the magnetic susceptibility\cite{Koba} 
showed a steep decrease without anisotropy 
by lowering the temperature in the 
compounds that exhibit an insulating ground state, 
this phase, which is next to the 
superconducting phase, is considered to be a nonmagnetic insulating (NMI) phase. 
We can see that 
the general feature of this phase diagram is analogous to the 
proposed phase diagram of $\kappa$-(ET)$_2$X, \cite{Kanoda} 
where the 10-K superconducting phase is situated near the 
Mott antiferromagnetic insulating (AFI) phase,\cite{Kino}
if the NMI phase in the present case is replaced by the AFI phase. 
 
\begin{figure}
\begin{center}
\epsfile{file=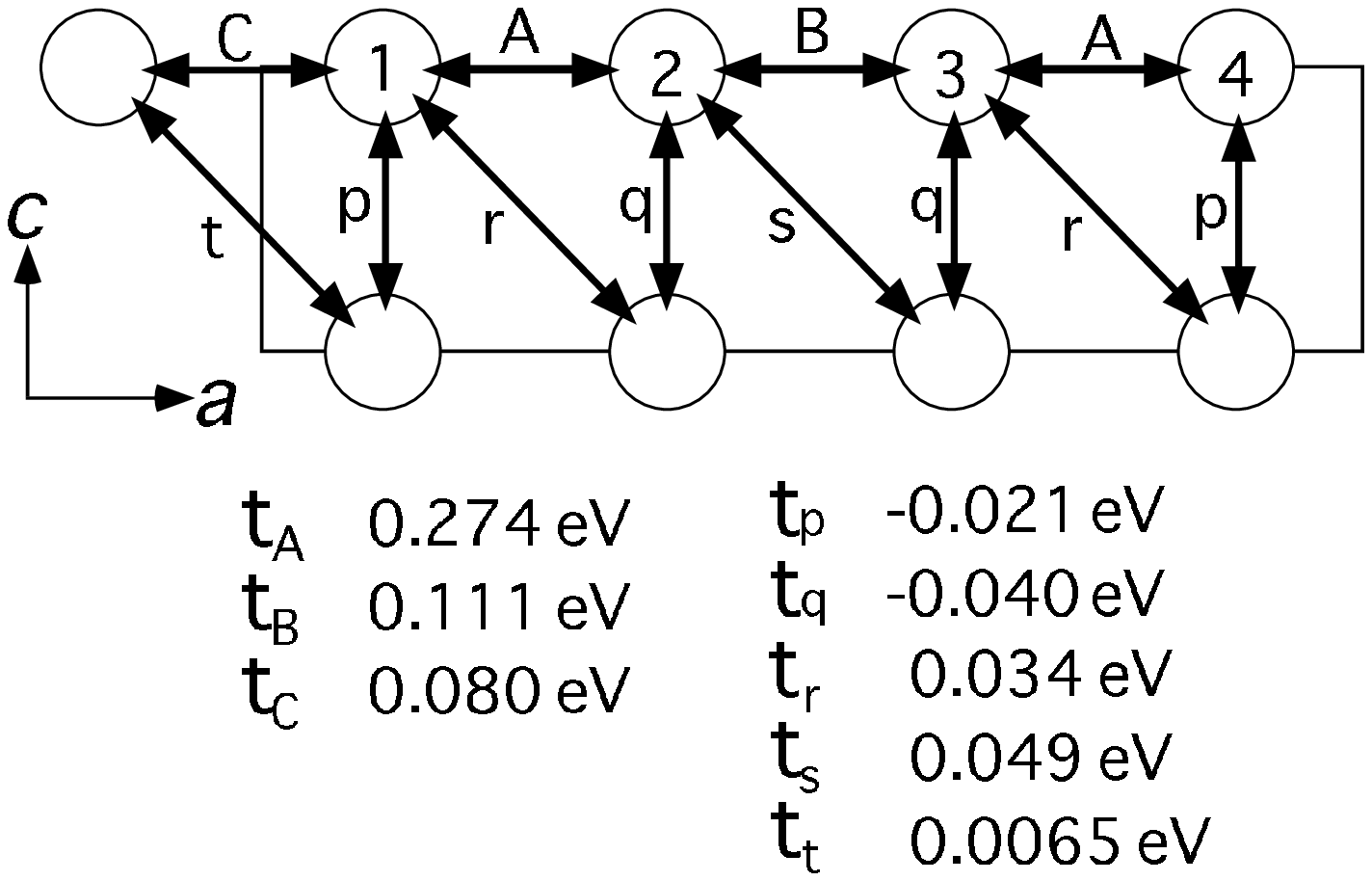,height=5cm}
\end{center}
\caption{A schematic view of the structure of the donor plane in 
$\lambda$-(BETS)$_2$GaX$_z$Y$_{4-z}$. 
The values of transfer integrals for Y=Cl, $z=0$
are given by $t=Es$, where $s$ is the value of the overlap integral 
from ref. 1 and $E=-10eV$.}
\label{model}
\end{figure}

$\lambda$-(BETS)$_2$GaX$_z$Y$_{4-z}$ consists of four BETS molecules 
in a unit cell and take layered structure of 2D BETS network, 
whose structure in the donor plane is 
schematically shown in Fig. \ref{model}. 
It is seen in Fig. \ref{model} that there exist  
dimers (1--2) and (3--4) which are symmetrically equivalent 
since their intradimer transfer integral $t_A$ is twice as large as 
the others. 
We note that values of $t_B$ and $t_C$ differ due to the crystal structure. 
BETS is electrically expressed as BETS$^{+1/2}$ in this family, 
and then there are two holes in a unit cell. 
The extended H\"uckel band calculations\cite{Koba} 
predict that these $\lambda$-type
compounds are metallic with a 2D Fermi surface 
very similar to those of the $\kappa$-type conductors, despite  
their large differences in the molecular arrangement.
The partially-filled higher two bands 
and the fully-occupied lower two bands are separated so  
we can consider the system effectively half-filled, 
thus strong correlation can lead the system to a Mott insulator. 
Since the behavior in the magnetic susceptibility in
the insulating state is reminiscent of that in the quasi-one-dimensional organic 
spin-Peierls compounds such as 
(TMTTF)$_2$PF$_6$\cite{Jerome} which also has 
an effectively half-filled band, 
we may speculate that the NMI phase is due to the spin-Peierls transition  
which is characteristic of one-dimensional materials, 
yet the 2D character of these BETS compounds may contradict this conjecture. 
Our aim in this paper is to clarify the origin of this NMI phase.  

In order to study the electronic structure of these BETS salts, 
we only consider their 2D donor plane
(Fig. \ref{model}) which consists of 
transfer integrals between BETS molecules and the on-site Coulomb interaction $U$
on BETS molecule. 
The Hamiltonian is written as

\begin{eqnarray}
H=\sum_{<i,j>}\sum_{\sigma} \left( t_{i,j}a^{\dagger}_{i\sigma}%
    a_{j\sigma}+h.c.\right) +
    U\sum_{i}n_{i\uparrow}n_{i\downarrow},
\label{eqn:Hamil}
\end{eqnarray}
where $<i,j>$ denotes the neighbor site pair, 
$\sigma$ is a spin index which takes $\uparrow$ and $\downarrow$, 
$n_{i\sigma}$ and  $a^{\dagger}_{i\sigma}$ ($a_{i\sigma}$) denote 
the number operator and the creation (annihilation) operator for the 
electron of spin $\sigma$ at the $i$th site, respectively. 

We treat $U$ in the Hartree-Fock (HF) approximation, 
as in ref. 3, and
assume that 
the periodicity of the electron system along the $a$-direction is 
the same as that of the original lattice. 
As for the $c$-direction, we considered two cases; 
one with the same periodicity as that of the original lattice system, 
i.e. the paramagnetic or ferromagnetic solutions along the $c$-direction, 
or one two-fold that of the original lattice system, 
i.e. the antiferromagnetic solution along the $c$-direction. 
Thus, there are four or eight different sites in a unit cell and the 
HF Hamiltonian in $k$-space is given by
\begin{full} 
\begin{eqnarray}
H^{HF}=\sum_{k\sigma}\left(\begin{array}{cccc}
a_{1k\sigma}\\
a_{2k\sigma}\\
\vdots\\
a_{mk\sigma} \end{array}\right)^{\dagger}
\left[h_{0}+U\left(\begin{array}{cc}
\begin{array}{cc}n_{1\bar{\sigma}} & \\
 & n_{2\bar{\sigma}}
\end{array} & 0 \\
0 & \begin{array}{cc}\ddots & \\
 & n_{m\bar{\sigma}}
\end{array}
\end{array}\right)\right]
\left(\begin{array}{cccc}
a_{1k\sigma}\\
a_{2k\sigma}\\
\vdots\\
a_{mk\sigma} \end{array}\right),
\label{eqn:HFHamil}
\end{eqnarray}
\end{full}
where $m=4$ or $m=8$ according to the unit cell size, 
$\bar{\sigma}$ is opposite to $\sigma$ and $a_{\nu k\sigma}$
is the Fourier transform of $a_{i\sigma}$, i.e.
$a_{\nu{k}\sigma}=\frac 1{\sqrt{N_{\rm cell}}}\sum_{\alpha}
e^{iR_{\alpha\nu}k}a_{(\alpha\nu)\sigma}$ 
and $N_{\rm cell}$ is the total number of unit cells.
$\alpha$ and $\nu$ denote the cell number and 
the number of the molecule, respectively. The $h_0$ term is a matrix 
which comes from the first term of eq. (\ref{eqn:Hamil}) and is 
diagonalized to lead to almost the same dispersionas that given by 
the extended H\"uckel band calculation.\cite{Koba}

In eq. (\ref{eqn:HFHamil}), the electron densities for each spin, $n_{\nu\sigma}$ 
($\nu=1,2,\cdots,m$ and $\sigma=\uparrow,\downarrow$), are determined 
self-consistently by 
\begin{equation}
n_{\nu\sigma}=\frac 1{N_{\rm cell}}\sum_{k}
\left<a_{\nu{k}\sigma}^{\dagger}a_{\nu{k}\sigma}\right>.
\end{equation}
They are calculated together with the equation for the Fermi energy 
implicitly given by
\begin{equation}
N_e=\sum_{\nu\sigma}n_{\nu\sigma},
\end{equation}
where $N_e$ is the total number of electrons in the unit cell in the 
highest occupied molecular orbital 
(HOMO) levels. ($N_e=6$ for m=4 or $N_e=12$ for m=8 in the system currently 
considered.)

\begin{figure}
\begin{center}
\epsfile{file=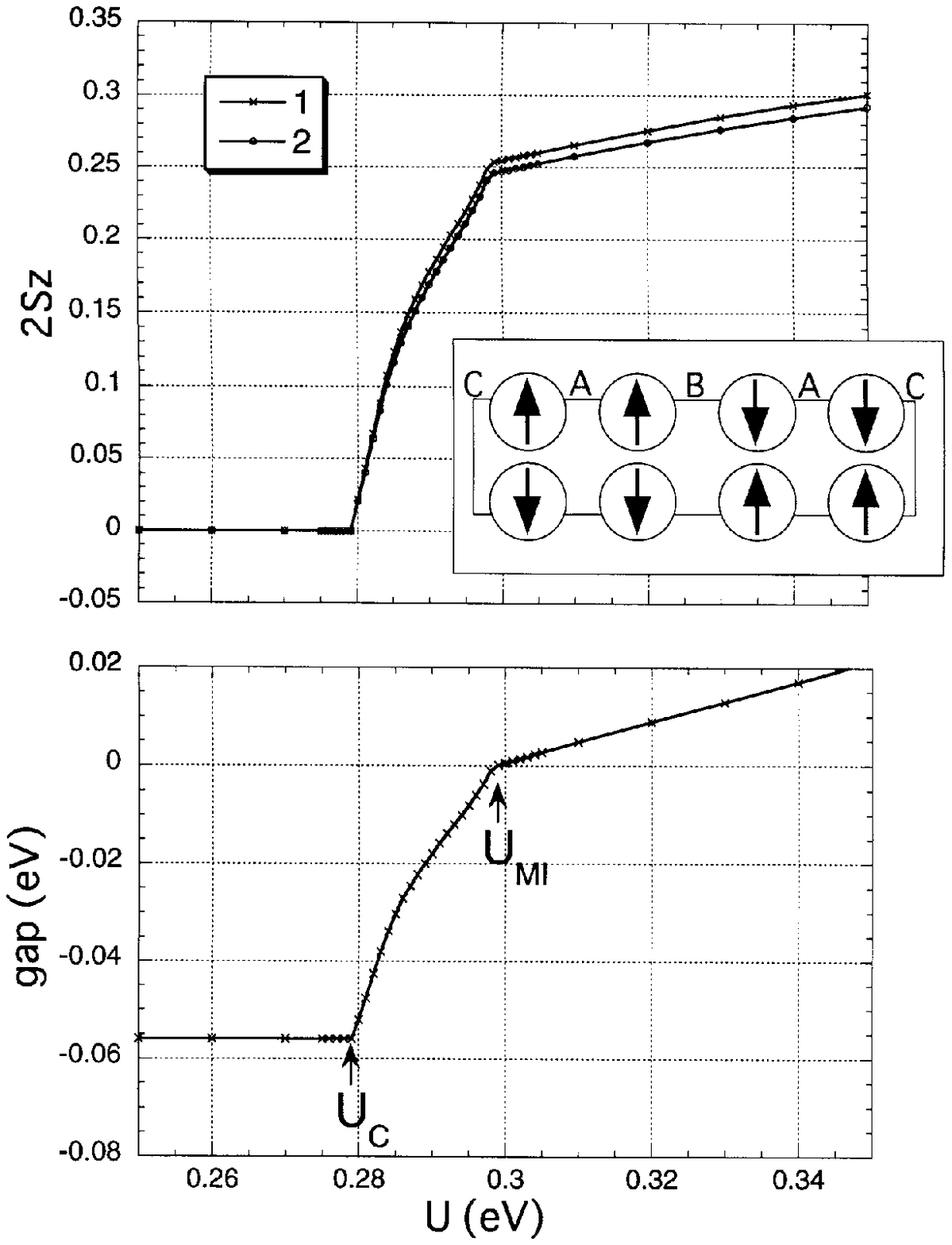,height=11.5cm}
\end{center}
\caption{U dependence of the absolute magnitude of spin moments, $S_z$, 
and the band gap (negative value implies a band overlap). }
\label{Sz}
\end{figure}

The calculations indicate that an antiferromagnetic (AF) order appears 
when $U$ exceeds some critical value and it affects the degree of 
the band overlap. 
The calculated absolute magnitude of spin moment, $S_z$, 
for molecules 1 and 2 and the band gap as a function of $U$ 
are shown in Fig. \ref{Sz}. 
We use the term `band gap' as a value of the lowest amount of energy for  
the highest (second) partially or fully filled band 
minus the highest amount of energy for the second (third) partially or fully filled 
band for m=4 (m=8), 
so a negative band gap implies a finite band overlap. 
The results indicate that there exist two phase transitions; 
a 2nd-order transition between a paramagnetic metal state and an 
AF metal (AFM) state at $U=U_c\simeq0.279$ eV and 
another 2nd-order transition from AFM state to AFI state at 
$U=U_{MI}\simeq0.296$ eV. 
The hole density at each site is slightly different from +0.5 
but is similar. 
In the AFM and AFI phases, the configuration of spin alignments 
is shown in the inset of Fig. \ref{Sz}. 
The direction of the spin moments 
inside the dimers are the same and 
a 2D AF ordering occurs between dimers. 
We note that $S_z(1)=-S_z(4)$, $S_z(2)=-S_z(3)$ 
in each row along the $a$-direction, 
and that $S_z(1)$ and $S_z(2)$ differ by a few \% mainly due to the differences of 
the transfer integrals $t_B$ and $t_C$. 

\begin{figure}
\begin{center}
\epsfile{file=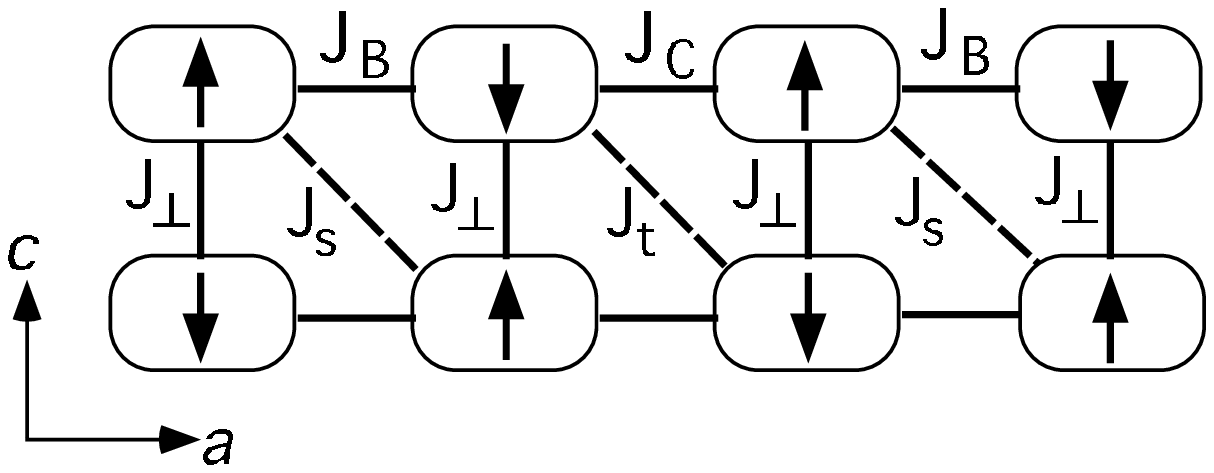,height=3cm}
\end{center}
\caption{Effective 2D Heisenberg model for localized spins in the dimer model
for $\lambda$-(BETS)$_2$GaBr$_x$Cl$_{4-x}$.}
\label{dimer}
\end{figure}

One can extract the essence 
by taking each dimer as a unit together with the effective 
overlap integrals between dimers. 
This view is called the dimer model in the case of 
$\kappa$-(ET)$_2$X.\cite{Kanoda,Kino} 
The above results show that in the large $U$ region, 
where the BETS compounds are considered to be located, 
the system can be viewed as a 2D localized spin system, 
therefore, we can construct an effective 2D Heisenberg spin model  
as schematically shown in 
Fig. \ref{dimer}. 
We estimate the superexchange coupling constants, $J$, 
using the relation $J\simeq 4t^2/U_{dimer}$
(Note that $U_{dimer}$ is the effective Coulomb interaction 
between two holes in a dimer), and conclude that
the magnitudes of the interactions between the spins are 
$J_C\simeq 0.52J_B, J_{\perp}\simeq 0.72J_B, J_s\simeq 0.19J_B$ and $J_t\simeq 0.02J_B$.
If we neglect $J_s$ and $J_t$, which are noticeably smaller than the others, 
the system can be viewed as dimerized AF Heisenberg chains
along the $a$-direction coupled by interchain coupling $J_{\perp}$
by which the Hamiltonian is described as 
\begin{eqnarray}
H=&J_B& \sum_{i\ even}{\mib S}_i\cdot{\mib S}_{i+\hat{a}}
    +J_C \sum_{i\ odd}{\mib S}_i\cdot{\mib S}_{i+\hat{a}} \nonumber\\
    &+&J_{\perp} \sum_{i}{\mib S}_i\cdot{\mib S}_{i+\hat{c}}, 
\label{eqn:spin}
\end{eqnarray} 
where ${\mib S}_i$ is the $S=1/2$ spin operator 
on the $i$th site and $\hat{a}$ and $\hat{c}$ denote 
the unit vector in the $a$-direction and the $c$-direction, respectively. 
 
In 2D quantum spin systems, quantum fluctuations, 
which we have not taken into account until now, 
result in the competition between 
the antiferromagnetic ordered state and 
the disordered spin liquid state which has a spin gap. 
In our model
there are two key parameters which determine which state will be realized;  
the degree of dimerization, i.e. the difference between $J_C$ and $J_B$, 
and the interchain coupling $J_{\perp}$. 
As a general tendency, 
the spin gap phase is stabilized in the region 
where the degree of dimerization is large enough. 
The interchain coupling $J_{\perp}$ prefers the AF state 
when $J_{\perp}\lsim J_C,J_B$. In our system, however,  
$J_{\perp}\sim J_C,J_B$ and it is not obvious which state is stabilized. 
Katoh and Imada\cite{Katoh} carried out quantum Monte Carlo simulations 
to the Hamiltonian (\ref{eqn:spin}), and determined the phase diagram 
in a parameter space of the magnitude of interchain coupling 
and the degree of dimerization 
(in their notation $J_B=J(1+\delta)$ and $J_C=J(1-\delta)$, 
$\delta$: degree of dimerization). 
According to their results, 
our spin system is located in the region close to the boundary 
between the AF phase and the spin gap phase, 
but slightly in the region of the AF state. 
It is expected that $J_s$, which has been neglected so far, 
pushes the system toward the spin gap phase. 
Considering the magnetic susceptibility experiments,\cite{Koba}
the NMI phase of $\lambda$-(BETS)$_2$GaX$_z$Y$_{4-z}$ 
is then thought to be in the spin gap regime. 

We can either see the system as 2-leg ladders along the $c$-direction 
whose interaction along the rungs is $J_B$ and 
that along the legs is $J_{\perp}$, 
coupled by interladder coupling $J_C$. 
Recently, a similar system, where two-leg ladders are coupled three-dimensionally,
was studied.\cite{Normand,Troyer} 
For example, in 
LaCuO$_{2.5},$\cite{Hiroi,Matsumoto,Kadono} 
the static susceptibility\cite{Hiroi} suggests
a spin liquid state with a spin gap, 
while NMR\cite{Matsumoto} and $\mu$SR\cite{Kadono} measurements indicate 
transition to a magnetically ordered phase.  
Theoritical studies\cite{Normand,Troyer} indicate that in a system located in 
the AF phase near the transition point to the 
spin liquid state,  
the static susceptibility has the form $\chi(T)=\chi_0+aT^2$, 
with a small $\chi_0$ which will not be identified easily by experiment. 
Therefore, we cannot exclude the possibility that 
$\lambda$-(BETS)$_2$GaX$_z$Y$_{4-z}$ 
is such an unconventional AF, 
close to the quantum critical point. 
For the moment we cannot decide beween the two possibilities, 
the 2D spin-gap phase or the AF close to criticality, 
and must wait for further experiments, such as 
NMR and $\mu$SR measurements for clarification. 

Increasing $J_C$ corresponds to the decrease of 
the degree of dimerization 
in the picture of dimerized Heisenberg chains, 
and to the increase of the interladder coupling in the picture of ladders system. 
In either picture, we see that, if $J_C$ increases, 
the system moves in the direction of the AF ordered state. 
Therefore, if one can synthesize compounds analogous to
$\lambda$-(BETS)$_2$GaX$_z$Y$_{4-z}$, 
but with a smaller degree of dimerization, 
the system should have AF ground state. 
By controlling the systems in such a way, 
we will be able to study quantum spin systems systematically. 

In summary, we have studied the nature of the nonmagnetic insulating 
phase in $\lambda$-(BETS)$_2$GaX$_z$Y$_{4-z}$. 
Hartree-Fock calculations suggest that the on-site Coulomb interaction $U$ 
causes an AF insulating state, 
which is a Mott insulator with one hole per dimer. 
Deducing from these results, 
we constructed an effective 2D localized dimer spin model 
with different exchange couplings. 
It is argued that quantum effects make this system a disordered spin liquid 
state with a spin gap, and we infer that
the system is located close to the quantum critical point.  

\acknowledgements
We thank A. Kobayashi, H. Tanaka and H. Kino 
for useful discussions and suggestions. 
This work was financially supported by a 
Grant-in-Aid for Scientific Research on Priority Area ``Anomalous Metallic
State near the Mott Transition'' (07237102) from the Ministry of Education, 
Science, Sports and Culture.

\end{document}